\documentclass[twocolumn,showpacs,preprintnumbers,amsmath,amssymb,superscriptaddress,floatfix,nofootinbib]{revtex4}

\usepackage{graphicx}
\usepackage{amsmath}
\usepackage{amsfonts}
\usepackage{amssymb}
\usepackage{color}
\usepackage{epsfig}
\usepackage{mathrsfs}
\usepackage{longtable,lscape}

\newcommand{\Slash}[1]{\ooalign{\hfil/\hfil\crcr$#1$}}

\begin{document}
\title{Role of $\Lambda(1670)$ in $\gamma p \to K^+ \eta \Lambda $ reaction near threshold}

\author{Li-Ye Xiao}
\affiliation{Department of Physics, Hunan normal University, Changsha, Hunan 410081, China}

\author{Qi-Fang L\"{u}}
\affiliation{Department of Physics, Zhengzhou University, Zhengzhou, Henan 450001, China}

\author{Ju-Jun Xie} \email{xiejujun@impcas.ac.cn}
\affiliation{Institute of Modern Physics, Chinese Academy of
Sciences, Lanzhou 730000, China} \affiliation{Research Center for
Hadron and CSR Physics, Institute of Modern Physics of CAS and
Lanzhou University, Lanzhou 730000, China} \affiliation{State Key
Laboratory of Theoretical Physics, Institute of Theoretical Physics,
Chinese Academy of Sciences, Beijing 100190, China}

\author{Xian-Hui Zhong}
\affiliation{Department of Physics, Hunan normal University, Changsha, Hunan 410081, China}

\begin{abstract}

The role of the $\Lambda(1670)$ resonance in the $\gamma p \to K^+
\eta \Lambda$ reaction near threshold is studied within an effective
Lagrangian approach. We perform a calculation for the total and
differential cross section of the $\gamma p \to K^+ \eta \Lambda$
reaction by including the contributions from the $\Lambda(1670)$
intermediate state decaying into $\eta \Lambda$ dominated by $K^-$
and $K^{*-}$ mesons exchanges, the nucleon pole and $N^*(1535)$
resonance decaying into $K^+ \Lambda$ dominated by exchanges of
$\omega$ and $K^-$ mesons. Besides, the non-resonance process and
contact terms to keep the total scattering amplitude gauge invariant
are also considered. With our model parameters, the total cross
section of this reaction is of the order of $1$ nanobarn at photon
beam energy $E_{\gamma} \sim 2.5$ GeV. It is expected that our model
predictions could be tested by future experiments.

\end{abstract}

\date{\today}

 \maketitle

\section{Introduction}{\label{introduction}}

One major goal of hadronic physics is to get the properties of
baryon resonances. There exists a huge amount of data especially
from $\pi N$ reactions for studying the nucleon and $\Delta(1232)$
resonances. The properties of most of these resonances are reliably
extracted by analyzing various experimental
data~\cite{Agashe:2014kda}, which also can be reasonably described
by the constituent quark models~\cite{Isgur:1978xj,
Capstick:1986bm,Loring:2001kx}. However, the situation changes when
we look at the properties of $\Lambda$ resonances. Large
uncertainties exist due to poor statistic of data and limited
knowledge of background contributions. Even for the well-established
low-lying negative parity states, such as $\Lambda(1405)$,
$\Lambda(1520)$, and $\Lambda(1670)$, their properties are still
controversial~\cite{Klempt:2009pi}, although they are four-star
ranking in the review of particle physics~\cite{Agashe:2014kda}.

To uncover the puzzles in the $\Lambda$ resonances, the
$K^-$-induced reactions provide us an important tool, especially the
reaction of $K^-p \to \eta \Lambda$. This reaction provides us a
clear place to study the low-lying $\Lambda$ resonances because only
the $\Lambda$ resonance contribute here due to the isospin selection
rule. Thus, when some accurate data on the differential and total
cross sections of the $K^-p \to \eta \Lambda$ reaction were reported
by the Crystal Ball Collaboration~\cite{Starostin:2001zz}, they were
analyzed with various theoretical models at once, e.g., effective
Lagrangian model~\cite{Liu:2012ge,Liu:2011sw}, chiral quark
model~\cite{Xiao:2013hca}. Recently, a comprehensive analysis of the
$K^- p$ scattering data of total and differential cross sections and
recoil polarizations was performed with a dynamical coupled-channel
model~\cite{Kamano:2014zba}. All of those theoretical analysis shows
that the $\Lambda(1670)$ dominates the $K^-p \to \eta \Lambda$
reaction around threshold for its strong coupling to $\eta \Lambda$,
while the other hyperon resonances give minor contributions.

In the present work, basing on the knowledge of the previous study
of the $K^-p \to \eta \Lambda$ reaction, we continue to study those
$\Lambda$ resonances from the $\gamma p\to  K^+ \eta \Lambda$
process. This process is also an important tool to gain information
on hadron resonance properties~\cite{Hanhart:2003pg,Zou:2009wp}. In
the $\gamma p\to  K^+ \eta \Lambda$ process, the contributions of
the $\Lambda$ resonance diagrams are considered to be caused by the
$K^-$ and $K^{*-}$ mesons exchanges between the initial photon and
proton. While the $\eta \Lambda$ production proceeds via the
excitation of the intermediate $\Lambda(1670)$~\footnote{It is worth
to mention that within the coupled channel chiral perturbation
theory, the $\Lambda(1670)$ resonance is dynamically generated from
the meson-baryon chiral interactions~\cite{GarciaRecio:2002td},
which indicates that the nature of $\Lambda(1670)$ could be
meson-baryon molecular state. However, the structure of the
$\Lambda(1670)$ resonance is not the purpose of the present work.
Hence, we will take it as an elementary particle considering its
finite decay width in its propagator.} resonance which has strong
coupling to the $\eta \Lambda$ channel.

The article is organized as follows. In the next section, the
formalism and ingredients necessary for our estimations are
presented, then we show our numerical results and discussions in
Sect. III. Finally, a short summary is given in Sec. IV.

\section{Formalism and ingredients} {\label{formalism}}

\begin{figure*}[htbp]
\begin{center}
\includegraphics[scale=0.4]{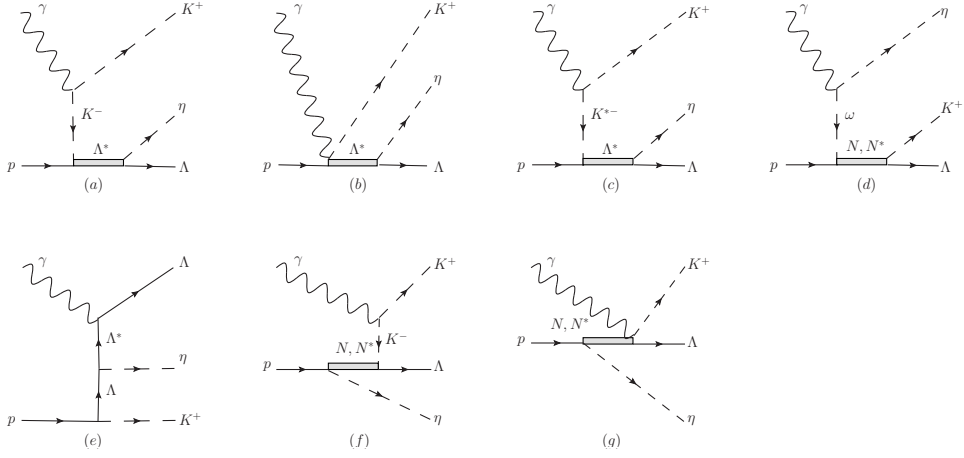}
\caption{Feynman diagrams for the $\gamma p \to K^+\eta\Lambda$
reaction.} \label{diagram}
\end{center}
\end{figure*}

We study the $\gamma p \to K^+ \eta \Lambda$ reaction within an
effective Lagrangian approach and the isobar model, which has been
successfully used in our previous
work~\cite{Xie:2007vs,Xie:2007qt,Xie:2013mua,Xie:2013wfa,Wu:2014yca,Xie:2014zga,Xie:2014kja}.
The basic Feynman diagrams for this process are depicted in
Fig.~\ref{diagram}. Since $\Lambda(1670)$ is close to the
$\eta\Lambda$ threshold and has strong coupling to $\eta \Lambda$,
we pay attention to the contributions from the $\Lambda(1670)$
($\equiv \Lambda^*$) resonance coupling to the $\eta \Lambda$ final
states. We also consider the contribution from $N(1535)$ ($\equiv
N^*$)~\footnote{We ignore the contributions from other low-lying
s-wave nucleon resonances which have small contributions to the
present calculation according to the Moorhouse selection
rules~\cite{Moorhouse:1966jn}, which pointed that those nucleon
resonances have very weak or vanish $\gamma NN^*$ coupling and hence
have weak $\omega NN^*$ coupling.} resonance which is caused by
$\omega$ exchange and decays into $K^+ \Lambda$ pair. It should be
emphasized that the contact terms are required to keep the gauge
invariant of the full scattering amplitude. In the calculations, we
have ignored some diagrams, such as the diagrams of the $\gamma p
\to p ({\rm or} ~ N^*) \to K^+ \Lambda ({\rm or} ~ \Lambda^*) \to
K^+ \eta \Lambda$ processes. The reason is that the couplings of the
interaction vertexes involved in these diagrams are weaker than
those diagrams we consider in Fig.~\ref{diagram} and the information
of the interaction vertexes involved in these ignored diagrams is
scarce.

To compute the contributions of the Feynman diagrams shown in
Fig.~\ref{diagram}, we adopt the interaction Lagrangian densities as
used in Refs.~\cite{Chiang:2002ah,Liu:2012kh}:
\begin{eqnarray}
{\cal L}_{\gamma K K} &=&
e(K^-\partial^{\mu}K^+-K^+\partial^{\mu}K^-)A_\mu, \\
{\cal L}_{\gamma K K^*}  &=& \frac{e g_{\gamma KK^*}}{m_{K^*}}
\varepsilon^{\mu\nu\alpha\beta}\partial_{\mu}K^*_\nu\partial_{\alpha}A_\beta
K, \\
{\cal L}_{\gamma \eta \omega} & = & \frac{e g_{\gamma
\eta\omega}}{m_{\omega}}\varepsilon^{\mu\nu\alpha\beta}\partial_{\mu}\omega_\nu\partial_{\alpha}A_\beta
\eta,
\end{eqnarray}
\begin{eqnarray}
{\cal L}_{\gamma \Lambda \Lambda^*}  &=& \frac{e \kappa_{\gamma
\Lambda \Lambda^*}}{2(m_{\Lambda} + m_{\Lambda^*})} \bar{\Lambda}^*
\gamma_5 \sigma_{\mu\nu} \Lambda F^{\mu \nu} + {\rm h.c.}, \\
{\cal L}_{K N \Lambda^*}  &=& g_{\bar{K} N \Lambda^*} \bar N K
\Lambda^* + {\rm h.c.}, \label{knlams} \\
{\cal L}_{K \Lambda N^*}  &=& g_{K \Lambda N^*} \bar \Lambda K N^* +
{\rm h.c.}, \label{knlams} \\
{\cal L}_{\eta \Lambda \Lambda^*}  & =& g_{\eta \Lambda \Lambda^*} \bar\Lambda \eta \Lambda^* + {\rm h.c.}, \\
{\cal L}_{\eta N N^*} &=& g_{\eta N N^*} \bar N \eta N^* + {\rm h.c.}, \\
{\cal L}_{K^* N \Lambda^*} &=& g_{\bar{K}^* N \Lambda^*} \bar N \gamma_5\gamma^{\mu} K^*_{\mu}\Lambda^*  + {\rm h.c.}, \\
{\cal L}_{\omega N N} &=& g_{\omega NN} \bar N (\gamma_{\mu} + \frac{\kappa_{\omega NN}}{2m_N}\sigma_{\mu\nu}\partial^\nu)\omega^{\mu} N , \\
{\cal L}_{\omega N N^*} &=& g_{\omega NN^*} \bar N \gamma_5\gamma^{\mu}\omega_{\mu}N^*  +  {\rm h.c.}, \\
{\cal L}_{K \Lambda N} & = & - \frac{g_{K\Lambda N}}{m_N+m_\Lambda}
\bar\Lambda \gamma_5 \gamma_{\mu} \partial^{\mu} K N, \label{pinpv} \\
{\cal L}_{\eta N N} & = & - \frac{g_{\eta N N}}{2m_N} \bar N
\gamma_5 \gamma_{\mu} \partial^{\mu} \eta N, \label{pinpv}
\end{eqnarray}
where $e = \sqrt{4\pi \alpha}$ ($\alpha$ is the fine-structure
constant), $A_{\mu}$ and $F_{\mu \nu}$ ($=
\partial_{\mu} A_{\nu} - \partial_{\nu}A_{\mu}$) are the photon field
and electromagnetic field tensor, respectively. We take
$\kappa_{\omega NN}$=0 as used in Ref.~\cite{Riska:2000gd} and
$\kappa_{\gamma N N^*} = 2.06$ obtained from the partial decay width
of $N^* \to N \gamma$. For $\kappa_{\gamma \Lambda \Lambda^*}$, we
determine it with $SU(3)$ flavor symmetry prediction, which gives
$\kappa_{\gamma \Lambda \Lambda^*} = 1.03$.~\footnote{Obtained with
the relation $\kappa_{\gamma N N^*}/\kappa_{\gamma \Lambda\Lambda^*}
= 2$ which was predicted by a chiral quark
model~\cite{Zhao:1998fn}.} For the coupling value $g_{\omega N N}$,
with the ratio of $g_{\omega N N}/g_{\rho N
N}=3.0$~\cite{Riska:2000gd}, and the value of $g_{\rho NN}\simeq
3.36$~\cite{Xie:2007vs}, we get $g_{\omega N N} \simeq 10.09$. For
the coupling constants $g_{N^* N \omega}$ and $g_{\Lambda^* N {\bar
K}^*}$, they are also obtained with the $SU(3)$ flavor symmetry
relation, which gives $g_{N^*N\omega} \simeq 0.43$ and $g_{\Lambda^*
N {\bar K}^*} \simeq 0.75$.~\footnote{Obtained with $g_{N^* N \rho}
=0.87$~\cite{Xie:2007qt}, $g_{N^* N \rho}/g_{N^* N \omega} = 2$ and
$g_{\Lambda^* N {\bar K}^*}/g_{N^* N \omega} =\sqrt{3}$, which is
evaluated from the vector meson-quark couplings using a chiral quark
model~\cite{Zhao:1998fn}.} Moreover, we take $g_{\eta NN} = 2.02$
and $g_{K \Lambda N} =-13.98$ as used in Ref.~\cite{Xie:2013wfa},
while the other coupling constants are determined from the partial
decay width of $K^*$, $\omega$, $\Lambda(1670)$, and $N(1535)$ as
listed in Tab~\ref{tabcc}.

\begin{table}[htbp]
\begin{center}
\caption{\label{tabcc} Parameters used in the present calculations.
[FS] means that the corresponding parameters are obtained from the
$SU(3)$ flavor symmetry.}
\begin{tabular}{ccccc}
\hline
State       & Width  &  Decay   & Adopted             & $g^2/4\pi$ \\
            & (MeV)  &  channel  & branching ratio     &  \\
\hline
$K^*$       & $51$   & $\gamma K$   & $9.9\times10^{-6}$  &$2.24\times 10^{-2}$\\
$\omega$    & $8$   & $\gamma \eta$ & $4.6\times10^{-6}$  &$9.75\times 10^{-3}$\\
$\Lambda(1670)$ & $35$ & $ \bar{K} N $     & $0.25$ & $9.20\times 10^{-3}$\\
$$                                 && $\Lambda\eta$  & 0.18 &$6.59\times 10^{-2}$\\
$$                                 &&$\bar{K}^* N$           & $-$  &$4.48\times 10^{-2}$[FS]\\
$$                           & &$\gamma \Lambda$ & $-$ &$8.44\times 10^{-2}$[FS]\\
$N(1535)$ & $150$ & $N\eta$       & 0.42 &$0.28$\\
$$                           && $\Lambda K$ & $-$ &$6.88\times 10^{-2}$~\cite{Liu:2012kh}\\
$$                           && $N \omega$ & $-$ &$1.50\times 10^{-2}$[FS]\\
\hline
\end{tabular}
\end{center}
\end{table}

The propagators of nucleon (or $\Lambda$ hyperon), vector mesons
$\omega$ and $K^{*-}$, and pseudo-scalar $K^-$ meson can be written in the forms
of
\begin{eqnarray}
G_{N/\Lambda}(p_{N/\Lambda}) &=& i \frac{\Slash{p}_{N/\Lambda} + m_{N/\Lambda}}{p^2_{N/\Lambda} - m^2_{N/\Lambda}}, \\
G^{\mu\nu}_{V}(p_V) &=&
-i \frac{g^{\mu\nu}-p^{\mu}_Vp^{\nu}_V/m_V^2}{p_V^2-m_V^2}, \\
G_{K}(p_K) &=& \frac{i}{p_K^2-m^2_K},
\end{eqnarray}
where $p_{N/\Lambda}$, $p_V$, and $p_K$ are the four-momentum of the
exchanged nucleon (or $\Lambda$ hyperon), vector mesons $\omega$ and
$K^{*-}$, and pseudo-scalar $K^-$ meson, respectively.

In addition, the propagators for $\Lambda(1670)$ and $N(1535)$
can be written in a Breit-Wigner form~\cite{Liang:2002tk},
\begin{eqnarray}
G_{R}(p_{R})=\frac{i(\Slash p_R + M_{R})}{p^2_{R}-M^2_{R}+i
M_{R}\Gamma_{R}},\label{G1405}
\end{eqnarray}
where $p_R$, $M_R$, and $\Gamma_{R}$ are the four-momentum, mass and
the total decay width of $\Lambda(1670)$ or $N(1535)$, respectively.

Besides, we need to include the form factors because the hadrons are
not point-like particles. We adopt here the common scheme used in
many previous works. In our calculations, the form factors for $K$,
$K^*$, $\omega$, off-shell nucleon pole, $N(1535)$, off-shell
$\Lambda$ pole, and $\Lambda^*$ are adopted the form advocated in
Refs.~\cite{Wu:2014yca,Feuster:1997pq,Penner:2002ma,Shklyar:2005xg},
\begin{eqnarray}
F(p^{2}_{ex},M_{ex}) = \left[ \frac{\Lambda_c^4}{\Lambda_c
^4+(p^{2}_{ex} - M^2_{ex})^2} \right ]^n,\label{form}
\end{eqnarray}
with $p_{ex}$ and $M_{ex}$ the four-momentum and mass of exchanged
hadron, respectively, $\Lambda_c$ the so-called cutoff parameter and
$n =1$ or $2$ depending on the specific
coupling~\cite{Machleidt:1987hj}. In the calculation, we adopt $n=1$
for $K$, $N(1535)$, and $\Lambda(1670)$, while $n=2$ for $K^*$,
$\omega$, nucleon pole, and $\Lambda$ pole. Besides, to minimize the
number of free parameters, we use the same cut off parameters
$\Lambda_c = 1.5$ GeV for all the exchanged hadrons for simplicity.

With those established effective Lagranians, propagators, and
coupling constants, the invariant scattering amplitudes for the
$\gamma p \rightarrow K^+ \eta\Lambda$ reaction can be obtained
straightforwardly by following the standard Feynman rules. The
amplitudes for Fig.~\ref{diagram}, can be written as
\begin{widetext}
\begin{eqnarray}
\mathcal{M}_a &=& e g_{\eta \Lambda \Lambda^*}g_{\bar{K} N \Lambda^*} F(p_{K}^2, m_K) F(p_{\Lambda^*}^2, m_{\Lambda^*}) \bar{u}(p_5, s_5)
G_{\Lambda^*}(p_{\Lambda^*})u(p_2,s_2)G_{K}(p_{K})(p_3^{\mu}-p_{K}^{\mu}) \varepsilon_{\mu}(p_1,s_1), \\
\mathcal{M}_b &=& e g_{\eta \Lambda \Lambda^*}g_{\bar{K} N
\Lambda^*} F(p_{K}^2, m_K) F(p_{\Lambda^*}^2, m_{\Lambda^*})
\bar{u}(p_5,s_5)
G_{\Lambda^*}(p_{\Lambda^*})\frac{p^{\mu}_{\Lambda^*}}{p_1 \cdot
p_{\Lambda^*}} u(p_2,s_2) \varepsilon_{\mu}(p_1, s_1), \\
\mathcal{M}_c &=& \frac{e g_{\eta \Lambda \Lambda^*} g_{\bar{K}^* N
\Lambda^*} g_{\gamma K K^*} F(p_{K^*}^2, m_{K^*}) F(p_{\Lambda^*}^2,
m_{\Lambda^*})}{m_{K^*}}
\bar{u}(p_5,s_5)G_{\Lambda^*}(p_{\Lambda^*}) \gamma_5
(\gamma_{\lambda} - \frac{p_{\Lambda^*\lambda} \Slash
p_{\Lambda^*}}{p_{\Lambda^*}^2}) u(p_2,s_2) \nonumber \\
&& \times G_{K^*}^{\lambda \nu}(p_{K^*}) \epsilon_{\mu \nu \alpha
\beta} p_{K^*}^{\mu} p_1^{\alpha} \varepsilon^{\beta}(p_1,s_1), \\
\mathcal{M}^N_d &=& \frac{e g_{K \Lambda N} g_{\omega NN} g_{\gamma
\eta \omega} F(p_{\omega}^2,m_{\omega})
F(p_{N}^2,m_{N})}{(m_{5}+m_{N})m_{\omega}} \bar{u}(p_5,s_5)
\gamma_{5} \Slash p_3 G_{N}(p_{N}) \gamma_{\lambda} u(p_2,s_2)
G_{\omega}^{\lambda \nu} (p_{\omega}) \epsilon_{\mu \nu
\alpha \beta} p_{\omega}^{\mu} p_1^{\alpha} \varepsilon^{\beta}(p_1,s_1), \\
\mathcal{M}_d^{N^*} &=& \frac{e g_{K \Lambda N^*} g_{\omega N N^*}
g_{\gamma \eta \omega} F(p_{\omega}^2,m_{\omega})
F(p_{N^*}^2,m_{N^*})} {m_{\omega}}
\bar{u}(p_5,s_5) G_{N^*}(p_{N^*}) \gamma_5 (\gamma_{\lambda} - \frac{p_{N^*\lambda} \Slash p_{N^*}}{p_{N^*}^2})u(p_2,s_2) \nonumber \\
&& \times G_{\omega}^{\lambda \nu} (p_{\omega}) \epsilon_{\mu \nu
\alpha \beta} p_{\omega}^{\mu} p_1^{\alpha}
\varepsilon^{\beta}(p_1,s_1), \\
\mathcal{M}_e &=& \frac{e \kappa_{\gamma \Lambda \Lambda^*} g_{\eta
\Lambda \Lambda^*} g_{K N \Lambda} F(p_{\Lambda^*}^{\prime
2},m_{\Lambda^*})
F(p_{\Lambda}^2,m_{\Lambda})}{(m_5+m_{\Lambda^*})(m_2+m_{\Lambda})}
\bar{u}(p_5,s_5) \gamma_5 \gamma^{\mu} \varepsilon_{\mu}(p_1, s_1)
G_{\Lambda^*}(p_{\Lambda^*}^{\prime})G_{\Lambda}(p_{\Lambda})
\gamma_5 \Slash p_3 u(p_2,s_2), \\
\mathcal{M}_f^N &=& \frac{e g_{K \Lambda N} g_{\eta NN}F(p_{K}^2,
m_{K})F(p_{N}^2, m_{N})}{(m_5+m_{N})(m_2+m_{N})} \bar{u}(p_5,s_5)
\gamma_5 \Slash p_K G_{N}(p_N)\gamma_5\Slash p_4G_{K}(p_K)
(p_3^{\mu} - p_K^{\mu}) \varepsilon_{\mu}(p_1,s_1), \\
\mathcal{M}_f^{N^*} &=& e g_{K \Lambda N^*} g_{\eta N N^*}
F(p_{K}^2, m_{K})F(p_{N^*}^2, m_{N^*}) \bar{u}(p_5,s_5)
G_{N^*}(p_{N^*}) u(p_2,s_2)G_{K}(p_K) (p_3^{\mu} - p_K^{\mu})
\varepsilon_{\mu}(p_1,s_1),
\end{eqnarray}
\end{widetext}
\begin{widetext}
\begin{eqnarray}
\mathcal{M}_g^{N} &=& \frac{e g_{K\Lambda N} g_{\eta N N}
F(p_{K}^2,m_{K})F(p_{N}^2, m_{N})}{(m_5+m_N)(m_2+m_N)}
\bar{u}(p_5,s_5) \frac{\gamma_5(\Slash p_1-\Slash p_3)p_3^{\mu}}{p_1 \cdot p_3}G_{N}(p_{N})\gamma_5 \Slash p_4u(p_2,s_2) \varepsilon_{\mu}(p_1,s_1), \\
\mathcal{M}_g^{N^*} &=& eg_{K \Lambda N^*} g_{\eta N N^*}F(p_{K}^2,
m_{K})F(p_{N^*}^2, m_{N^*})\bar{u}(p_5,s_5) \frac{p_K^{\mu}}{p_1
\cdot p_K} G_{N^*}(p_{N^*}) u(p_2,s_2) \varepsilon_{\mu}(p_1,s_1),
\end{eqnarray}
\end{widetext}
where the sub-indices $a$, $b$, $c$, $d$, $e$, $f$, and $g$ stand
for the diagrams shown in Fig.~\ref{diagram}. The $p_1$, $p_2$,
$p_3$, $p_4$ and $p_5$ represent the four-momentums of the photon,
proton, $K^+$ meson, $\eta$ meson and $\Lambda$ hyperon,
respectively. The $s_1$, $s_2$ and $s_5$ are the spin projections of
the photon, proton and $\Lambda$ hyperon, respectively. $p_K$ and
$p_{K^*}$ ($=p_1-p_3$) correspond to the four-momentum of exchanged
$K^-$ and $K^{*-}$ meson, respectively. $p_{\omega}$ ($=p_1-p_4$) is
the four-momentum of exchanged $\omega$ meson. $p_{\Lambda}$
($=p_2-p_3$) is the four-momentum of exchanged $\Lambda$ hyperon.
$p_N$ and $p_{N^*}$ ($=p_2-p_4$) are the four-momentum of
intermediate nucleon and nucleon resonances, respectively.
$p_{\Lambda^*}$ ($=p_4+p_5$) (Fig.~\ref{diagram} (a), (b) and (c))
and $p^{\prime}_{\Lambda^*}$ ($=p_1-p_5$) (Fig.~\ref{diagram} (e))
are the four-momentum of the intermediate $\Lambda(1670)$ resonance.

The contact terms illustrated in Fig.~\ref{diagram} (b) and (g)
serve to keep the full amplitude (${\cal M}$) gauge invariant. By
including the amplitude ${\cal M}_b$ and ${\cal M}^{N,~N^*}_g$, it
is easy to show that the total amplitude satisfies the gauge
invariance
\begin{eqnarray}
p_1 \cdot {\cal M} = 0
\end{eqnarray}
with
\begin{eqnarray}
{\cal M} &=& {\cal M}_a + {\cal M}_b + {\cal M}_c + {\cal M}^N_d +
{\cal M}^{N^*}_d + {\cal M}_e \nonumber \\
&& + {\cal M}^N_f + {\cal M}^{N^*}_f + {\cal M}^N_g + {\cal
M}^{N^*}_g.
\end{eqnarray}

Finally, the cross section for $\gamma p \to K^+ \eta \Lambda$
reaction can be calculated with
\begin{eqnarray}
d\sigma (\gamma p\to  K^+ \eta\Lambda) &=& \frac{1}{8E_{\gamma}}
\sum_{s_1, s_2, s_5} |{\cal M}|^2\frac{d^{3} p_{3}}{2E_{3}} \frac{d^{3} p_4}{2E_4} \frac{m_{\Lambda}d^{3}p_{5}}{ E_{5}}\nonumber\\
&& \times\delta^4(p_{1}+p_{2}-p_{3}-p_{4}-p_{5}),
\label{eqcs}
\end{eqnarray}
where $E_3$, $E_4$, and $E_5$ are the energies of the final
particles $K^+$, $\eta$, and $\Lambda$ hyperon, respectively;
$E_{\gamma}$ is the energy of incident photon at laboratory frame.

\section{Numerical results and discussions}

In this section we show our theoretical results of the total and
differential cross sections of the $\gamma p \to K^+ \eta \Lambda$
reaction near $\eta$-meson production threshold.

\subsection{Total cross section}

\begin{figure}[htbp]
\begin{center}
\includegraphics[scale=0.7]{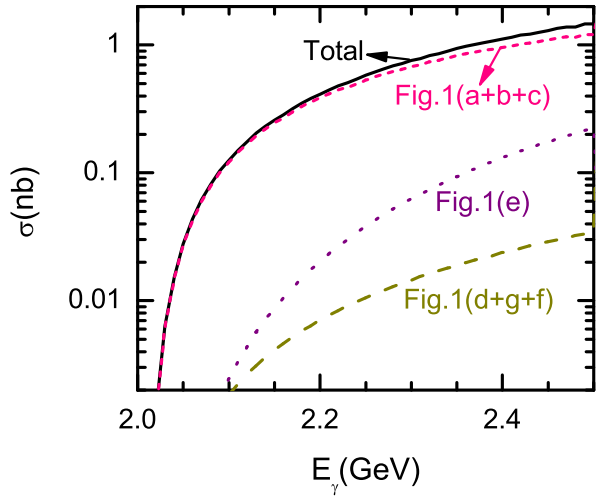}
\caption{(Color online) The cross sections vs photon beam energy
($E_{\gamma}$) for the $\gamma p \to K^+\eta\Lambda$ reaction from
present calculation obtained with the parameters in the
table~\ref{tabcc}. The solid curve represents the total cross
sections including all the contributions of Fig.~\ref{diagram}.
Different contributions of Fig.~\ref{diagram} (a+b+c),
Fig.~\ref{diagram} (e) and Fig.~\ref{diagram} (d+f+g) are indicated
explicitly by the legends in the figures.} \label{total}
\end{center}
\end{figure}

The total cross section versus excess energy $E_{\gamma}$ for the
$\gamma p \to K^+ \eta \Lambda$ reaction is calculated by using a
Monte Carlo multi-particle phase space integration program. In
Fig.~\ref{total}, we plot the total cross section as a function of
the photon beam energy $E_{\gamma}$ in the region of
$2.0<E_{\gamma}<2.5$ GeV. From Fig.~\ref{total}, one can clearly see
that the $\Lambda(1670)$ resonance [i.e., the contributions from the
diagrams (a), (b), and (c) in Fig.~\ref{diagram}] gives a dominant
contribution to the reaction from the threshold to $E_{\gamma} =
2.5$ GeV, while the contributions from the other diagrams shown in
Fig.~\ref{diagram} are small.

\begin{figure}[htbp]
\begin{center}
\includegraphics[scale=0.70]{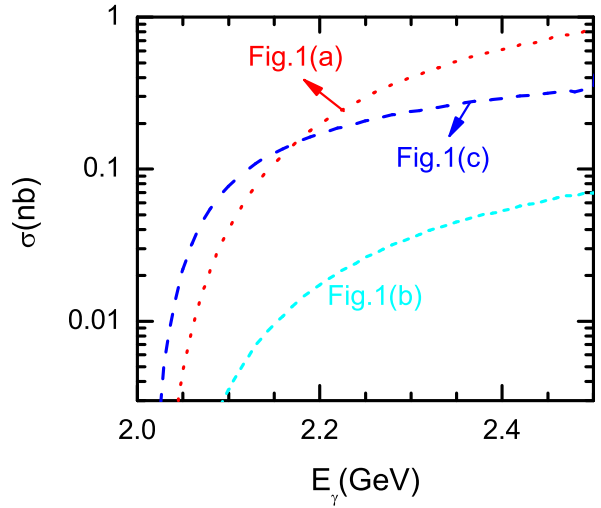}
\caption{(Color online) The cross sections vs photon beam energy
($E_{\gamma}$) for the $\gamma p \to K^+\eta\Lambda$ reaction from
$\Lambda(1670)$ resonance contribution. The contributions of $K^-$
meson exchange [Fig.~\ref{diagram} (a)], $K^{*-}$ vector-meson
exchange [Fig.~\ref{diagram} (c)] and contact term
[Fig.~\ref{diagram} (b)] are indicated with red dotted line, blue
dashed line and cyan short dashed line, respectively.}
\label{total1670}
\end{center}
\end{figure}

The contribution of the $\Lambda(1670)$ resonance includes three
parts: (i) $K^-$ meson exchange [Fig.~\ref{diagram} (a)]; (ii)
contact term [Fig.~\ref{diagram} (b)]; and (iii) $K^{*-}$ vector
meson exchange [Fig.~\ref{diagram} (c)]. The relative importance of
these three process to the $\gamma p \to K^+ \eta \Lambda$ reaction
is obviously shown in Fig.~\ref{total1670}. From which we can see
that the contribution from $K^{*-}$ vector meson exchange is larger
than that from $K^-$ meson exchange between the reaction threshold
and $E_{\gamma}\approx 2.16$ GeV. But, with increase of photon
energies $E_{\gamma}$, the contribution from $K^-$ meson exchange
grows faster than the one from $K^{*-}$ vector meson exchange. This
can be easily understood since the value of the coupling constant
$g^2_{\bar{K}^* N \Lambda(1670)}$ obtained according to the $SU(3)$
flavor symmtry is about $5$ times larger than $g^2_{\bar{K} N
\Lambda^*(1670)}$ determined from the partial decay width of
$\Lambda(1670) \to \bar{K} N$. Thus, the contribution from $K^{*-}$
vector meson exchange is dominant around the threshold. However,
with the photon energy increasing, the contribution of $K$-exchange
is to become more important for the faster increase of the
contribution from the $K$-exchange. The contribution from contact
term is small in the whole energy region considered in the present
work.

Meanwhile, due to the uncertainty of the form factors, it is
necessary to consider the effects of form factors on the cross
sections. For the dominant contributions of $\Lambda(1670)$ to the
reaction, we just need consider the influence of the form factors on
the cross section from the $K^{*-}$- and $K^-$-exchanges.

\begin{figure}[htbp]
\begin{center}
\includegraphics[scale=0.70]{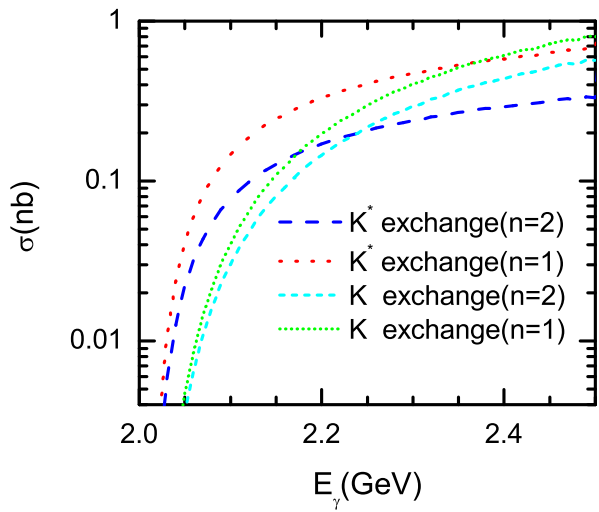}
\caption{(Color online) The cross sections vs photon beam energy
($E_{\gamma}$) for the $\gamma p \to K^+\eta\Lambda$ reaction from
$K^-$ and $K^{*-}$ meson-exchanges
 contributions with the different values of $n$ with cutoff parameter $\Lambda_c=1.5$ GeV.} \label{diffn}
\end{center}
\end{figure}

Firstly, we consider the form factors effects on the cross section
with different values of $n$ ($n$=1 or $n$=2) by fixing the cutoff
parameter with $\Lambda_c=1.5$ GeV. The results are shown in
Fig.~\ref{diffn}. We can see that the cross section is sensitive to
$n$. With $n=2$ the cross section is obviously suppressed compared
with that of $n=1$. The difference between the cross section
predicted with $n$=1 and that predicted with $n$=2 becomes more and
more large with the photon energy $E_{\gamma}$ increasing. For
example, at photon energy $E_{\gamma}=2.5$ GeV the cross section for
$K^-$ exchange varies from 0.5 nb to 0.8 nb with the values of $n$
changed from 2 to 1. While, for $K^{*-}$ exchange, the cross section
varies from 0.3 nb to 0.7 nb.

\begin{figure}[htbp]
\begin{center}
\includegraphics[scale=0.7]{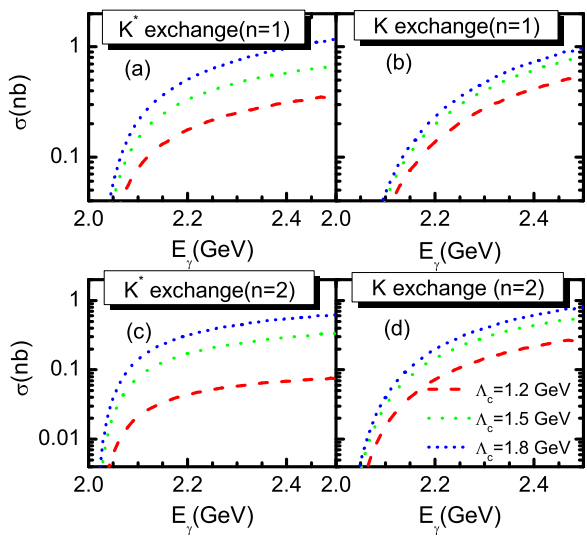}
\caption{(Color online) (a)-(b): the cross sections vs photon beam
energy ($E_{\gamma}$) for the $\gamma p \to K^+\eta\Lambda$ reaction
from the exchanged $K^-$ and $K^{*-}$ mesons contributions with the
different values of cutoff parameter $\Lambda_c$ with $n=1$;
(c)-(d): as in (a)-(b) but with $n=2$.} \label{diffcutoff}
\end{center}
\end{figure}

Next, we consider the form factors effects on the cross sections
with different cutoff parameters $\Lambda_c$.  In
Fig.~\ref{diffcutoff}, we show the predicted cross sections of the
$K^-$ and $K^{*-}$ exchanges with three typical cutoff parameters
$\Lambda_c=1.2, ~1.5, ~1.8$ GeV, respectively. From the figure, it
clearly see that the predicted cross sections have a strong
dependence on the cutoff parameter $\Lambda_c$. Considering a 20\%
uncertainty of the cutoff parameter, i.e., $\Lambda_c= 1.5 \pm 0.3$
GeV, we find that the uncertainty of the predicted cross sections
can reach to $100\%-200\%$.

Finally, it should be pointed out that our results might bear some
uncertainties from the coupling constant
$g_{\Lambda(1670)\bar{K}^{*}N}$, since it is estimated from the
$SU(3)$ flavor symmetry. However, there is no experimental data on
this reaction, we will leave those issues to further studies in the
future.

As a whole, with our model parameters, $\Lambda(1670)$ with
exchanged $K^-$ and $K^{*-}$ mesons gives the dominant contribution
to the $\gamma p \to K^+ \eta \Lambda$ reaction, while the
contributions from nucleon, $N(1535)$ resonance and $t$-channel are
negligibly small. The form factors have large effects on the total
cross section.

\subsection{Differential cross section}

\begin{figure}[htbp]
\begin{center}
\includegraphics[scale=0.45]{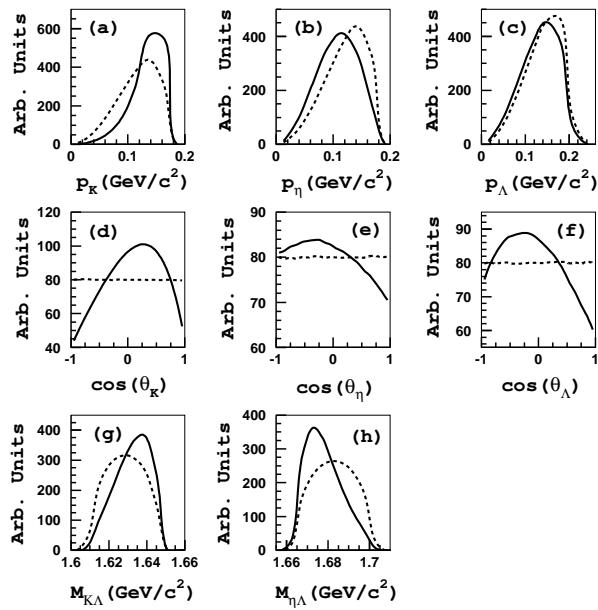}
\caption{Difference cross sections (solid lines) for the $\gamma p
\to \eta\Lambda  K^+$ at the excess energy $E_{\gamma}=2.1$ GeV and
phase-space distribution (dashed lines). (a)-(c): the momentum
distribution of outgoing $K^+$ meson, $\eta$ meson and hyperon
$\Lambda$, respectively; (d)-(f): the angular distribution of the
$K^+$ meson, $\eta$ meson and hyperon $\Lambda$ in the total
center-of-mass frame, respectively; (g)-(h): the invariant mass
spectrum of the outgoing hyperon $\Lambda$ and $K^+$ meson, outgoing
$\Lambda$ and $\eta$ meson, respectively.} \label{dif1}
\end{center}
\end{figure}

\begin{figure}[htbp]
\begin{center}
\includegraphics[scale=0.45]{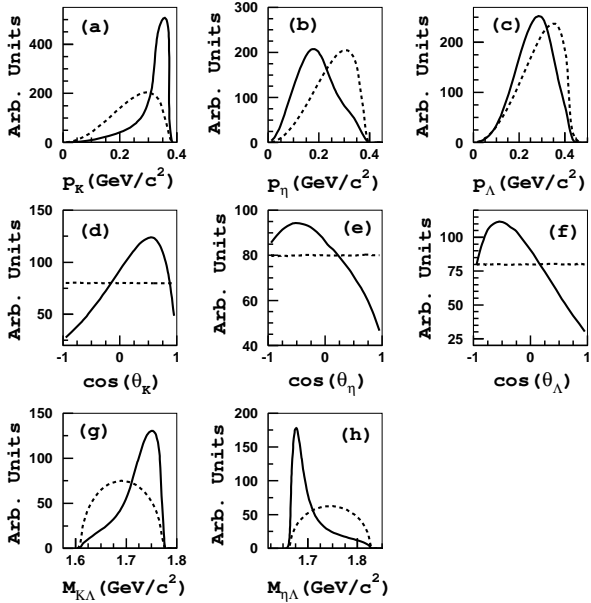}
\caption{As in Fig.~\ref{dif1} but for the case of $E_{\gamma} =
2.4$ GeV.} \label{dif2}
\end{center}
\end{figure}

In addition to the total cross section, we also study the
differential cross section for $\gamma p \to K^+ \eta \Lambda $
reaction. The corresponding momentum distribution and angular
distribution of outgoing $K^+$ meson, $\eta$ meson, and hyperon
$\Lambda$, the $K \Lambda$ and $\eta \Lambda$ invariant mass
spectrum in the center-of-mass frame at two energy points
$E_{\gamma}=2.1$, 2.4 GeV are shown in Fig.~\ref{dif1} and
Fig.~\ref{dif2}, respectively.

From the figures, we see that the $K^+$ meson and $\eta$ meson
momentum distributions are very different from the results with
phase space only. However, the $\Lambda$ hyperon momentum
distributions is similar to the results with phase space only.
Furthermore, from the angular distributions shown in the
figures, we see that $K^+$ meson has a large distribution at forward
angles. While the $\eta$ meson and hyperon $\Lambda$ have large
distributions at the backward angles. Finally, from the $\eta
\Lambda$ invariant mass spectrum, it is seen a obvious peak at
$M_{\eta\Lambda} \approx 1.675$ GeV, which is from the contribution
of $\Lambda(1670)$.

It should be pointed out that the effective Lagrangian approach is a
convenient tool to catch the qualitative features of the $\gamma
p\to  K^+ \eta \Lambda$ reaction; however, it is not consistent with
the unitary requirements. In principle, the unitary is important for
extracting the parameters of the baryon resonances from the
experimental data~\cite{Kamano:2009im,Suzuki:2009nj}, especially for
those processes involving many intermediate couple channels and
three-particle final states~\cite{Kamano:2008gr,Kamano:2011ih}.
Thus, the unitary might introduce effects on our model predictions.
Meanwhile, couple channel effects can not be taken into account in
our calculations. Our model calculation constitutes the first step
in this direction.

\section{Summary and Conclusions }

In this work, we investigate the total and differential cross
sections of the $\gamma p \to K^+ \eta \Lambda$ reaction within an
effective Lagrangian model. It is shown that the resonant diagrams
induced by $K^{*-}$ and $K^-$ mesons provide the most important
contributions. It is also found that the contributions from nucleon
pole and $N(1535)$ due to the $\omega$ meson exchange and $K$ meson
exchange are negligibly small. It should be remarked the form
factors for exchanged $K^{*-}$ vector meson and exchanged $K^-$
meson have a significant effect on the cross section.

We also studied the differential cross section at beam energy
$E_{\gamma}=2.1$, 2.4 GeV. It is found that $K^+$ meson has forward
angle distribution, while $\eta$ meson and hyperon $\Lambda$ have
backward angle distribution. We expect that future experiments will
provide a test to our model calculations.

\section*{Acknowledgments}

This work is partly supported by the National Natural Science
Foundation of China (Grants No. 11475227, No. 11075051 and No.
11375061), the Hunan Provincial Natural Science Foundation (Grant
No. 13JJ1018), and the Hunan Provincial Innovation Foundation for
Postgraduate.

\bibliographystyle{unsrt}

\end{document}